\documentclass[prl,twocolumn,noshowpacs,noshowkeys,preprintnumbers,floatfix,
  nofootinbib,superscriptaddress]{revtex4-1}
\usepackage{textcomp} %
\usepackage{amsfonts} 
\usepackage{amssymb} 
\usepackage{amsmath} 
\usepackage{subfigure} 
\usepackage{graphicx} 
\usepackage{array} 
\usepackage{dcolumn} 
\usepackage{bm} 
\usepackage{latexsym} 
\usepackage{longtable} 
\usepackage{hyperref} 
\graphicspath{{./figs/}}

\begin{document}


\title{Kaon $B$-parameter from improved
  staggered fermions in $N_f=2+1$ QCD }
\author{Taegil Bae}
\affiliation{
  Korea Institute of Science and Technology Information,
  Daejeon, 305-806, South Korea 
}
\author{Yong-Chull Jang}
\affiliation{
  Lattice Gauge Theory Research Center, FPRD, and CTP, \\
  Department of Physics and Astronomy,
  Seoul National University, Seoul, 151-747, South Korea
}
\author{Chulwoo Jung}
%
%
%
\affiliation{
  Physics Department, Brookhaven National Laboratory,
  Upton, NY11973, USA
}
\author{Hyung-Jin Kim}
\affiliation{
  Lattice Gauge Theory Research Center, FPRD, and CTP, \\
  Department of Physics and Astronomy,
  Seoul National University, Seoul, 151-747, South Korea
}
\author{Jangho Kim}
\affiliation{
  Lattice Gauge Theory Research Center, FPRD, and CTP, \\
  Department of Physics and Astronomy,
  Seoul National University, Seoul, 151-747, South Korea
}
\author{Jongjeong Kim}
\affiliation{
  Physics Department,
  University of Arizona,
  Tucson, AZ 85721, USA
}
\author{Kwangwoo Kim}
\affiliation{
  Lattice Gauge Theory Research Center, FPRD, and CTP, \\
  Department of Physics and Astronomy,
  Seoul National University, Seoul, 151-747, South Korea
}
\author{Sunghee Kim}
\affiliation{
  Lattice Gauge Theory Research Center, FPRD, and CTP, \\
  Department of Physics and Astronomy,
  Seoul National University, Seoul, 151-747, South Korea
}
\author{Weonjong Lee}
%
%
%
%
\affiliation{
  Lattice Gauge Theory Research Center, FPRD, and CTP, \\
  Department of Physics and Astronomy,
  Seoul National University, Seoul, 151-747, South Korea
}
\author{Stephen R. Sharpe}
%
%
%
\affiliation{
  Physics Department,
  University of Washington,
  Seattle, WA 98195-1560, USA
}
\author{Boram Yoon}
\affiliation{
  Lattice Gauge Theory Research Center, FPRD, and CTP, \\
  Department of Physics and Astronomy,
  Seoul National University, Seoul, 151-747, South Korea
}
\collaboration{SWME Collaboration}
\date{\today}
\begin{abstract}
We present a calculation of the kaon $B$-parameter, $B_K$,
using lattice QCD. 
We use improved staggered valence and sea fermions, 
the latter generated by the MILC collaboration with 
$N_f=2+1$ light flavors.
To control discretization errors, we use four different lattice
spacings ranging down to $a\approx 0.045\;$fm.
The chiral and continuum extrapolations are done
using SU(2) staggered chiral perturbation theory.
Our final result is 
$\hat{B}_K = 0.727 \pm 0.004 (\text{stat}) \pm 0.038 (\text{sys})$,
where the dominant systematic error is from our use of
truncated (one-loop) matching factors.
\end{abstract}
\maketitle
%
%
%

CP violation was first observed in the kaon system in 1964~\cite{FitchCronin},
but only in the last few years has
it been possible to use this classic result to learn about
the parameters of the Standard Model (SM).
CP violation in the SM is due to the phase in the Cabibbo-Kobayashi-Maskawa
(CKM) matrix, and leads to CP violation in kaon mixing---so-called
indirect CP violation---through Feynman diagrams 
involving virtual charm and top quarks.
Integrating out heavy quarks ($c$, $b$ and $t$) and the $W$ and $Z$-bosons,
one finds that a prediction for indirect CP violation requires the
calculation of the matrix element of a local four-fermion
operator between a $K_0$ and a $\overline K_0$.
To calculate this matrix element 
(which is parametrized by the kaon $B$-parameter, $B_K$)
requires control over the non-perturbative physics of the strong
interactions. The only known quantitative method to calculate such
matrix elements from first principles is lattice Quantum Chromodynamics 
(LQCD), and it is only very recently that lattice calculations
have begun to control all sources of error.

In this note we present a calculation of $B_K$ using improved
staggered fermions. Fully controlled results using other 
types of lattice fermion have been obtained previously, 
using valence domain-wall quarks with
staggered sea quarks~\cite{ALV-09},
using domain-wall valence and sea fermions~\cite{rbc-uk-08-1,rbc-uk-2010-1}.
and, very recently, using Wilson valence and sea fermions~\cite{BMW}.
We also presented a result using a partial data set in Ref.~\cite{wlee-10-3},
which we refer to as SW-1 in the following.
What sets our work apart from that using valence domain-wall fermions is that
we are able to use smaller values of the lattice spacing, $a$,
thus potentially providing better control over the continuum limit.
%
%
We also have very small statistical errors, such that the
final statistical error in $B_K$ is $\sim 0.5\%$.
In addition, our use of a different fermion
discretization provides a highly non-trivial
cross-check of the other results, analogous to the use of a different
experimental technique.

The required matrix element is parametrized by 
\begin{equation}
B_K(\mu,\text{R}) = 
\dfrac{\langle K^0 | O_{\Delta S = 2}(\mu,\text{R}) | \bar{K}^0 \rangle}
{8f_K^2 M_K^2/3}
\label{eq:bkdef}
\end{equation}
where $\text{R}$ is a specific regularization scheme chosen to
define the operator 
$O_{\Delta S = 2} = \sum_{\nu}
[ \bar{s} \gamma_\nu (1 - \gamma_5) d ]
[ \bar{s} \gamma_\nu (1 - \gamma_5) d ]$, 
and $\mu$ is the corresponding
renormalization scale.
We use lattice regularization, and then convert to the
continuum $\overline{\rm MS}$ scheme 
(using naive dimensional regularization for the $\gamma_5$)
using one-loop matching factors from Ref.~\cite{KLS}.
At the end we convert to the renormalization-scale
invariant quantity $\widehat{B}_K$.

We use improved staggered fermions for both valence and sea quarks.
The advantages of staggered fermions are that they are computationally
inexpensive and that they retain a remnant of chiral symmetry.
The latter property implies that the
matrix element in the numerator of
Eq.~(\ref{eq:bkdef}) vanishes when $M_K\to0$ because of the
``left-left'' chiral structure of $O_{\Delta S=2}$.
Without chiral symmetry, $O_{\Delta S=2}$ mixes with operators
with ``left-right'' structure, whose matrix elements do not
vanish when $M_K\to 0$, and are thus enhanced.
For both staggered and domain-wall fermions, such mixing is not
allowed. 
Mixing with chirally enhanced operators is allowed for Wilson fermions,
but appears now to be controllable~\cite{BMW}.

For the sea quarks, we use the MILC collaboration's
publicly available ensembles generated using
$2+1$ flavors of asqtad staggered fermions~\cite{milc-rmp-09}.
Here ``$2+1$'' indicates
degenerate up and down quarks and a heavier, nearly physical, strange quark. 
Each staggered lattice flavor describes four continuum fermions,
usually called ``tastes''. This unwanted degeneracy is
removed by the fourth-root prescription,
and we assume this leads to no problems with the continuum limit.
%

For the valence sector we choose HYP-smeared staggered fermions. 
%
%
We prefer these to asqtad fermions because they are more continuum-like,
e.g.
the breaking of taste symmetry is smaller by a factor of $\sim 3$%
~\cite{wlee-08-2}.
Nevertheless, taste-breaking induced by discretization errors
leads to significant complications in the analysis.
This enters both through mass splittings between 
kaons of different tastes and by inducing additional operator
mixing. Both these effects, as well as those due to the use of
different types of valence and sea quarks, and to the use of the
fourth-root prescription, can be incorporated into the chiral
effective theory describing staggered fermions---staggered
chiral perturbation theory (SChPT)~\cite{LS,AB03,VdWS,wlee-10-3}.
Specifically, we use SU(2) SChPT---in which only the up and
down quarks are treated as light.
After fitting to the forms predicted by SChPT, one can
then remove ``by hand'' the taste-breaking discretization errors.
%


%
We use the MILC asqtad lattices listed in Table~\ref{tab:milc-lat}
for the present work.
%
%
The most important changes since our previous work, SW-1, are
the addition of a fourth, finer, lattice spacing (the ``ultrafine''
ensemble U1) and the 9 or 10-fold increase in the number of
measurements on several of the ``coarse'' ensembles ($a\approx 0.12\;$fm) 
and also on the ``fine'' ($a\approx 0.09\;$fm) ensemble F1.
%
%
%
%
%
\begin{table}[tbp]
  \caption{MILC ensembles used to calculate $B_K$~\cite{milc-rmp-09}.
    $a$ is the nominal lattice spacing, $m_\ell$ ($m_s$) the light (strange)
    sea-quark mass, ``ens'' the number of gauge configurations 
    and ``meas'' the number of measurements per configuration. 
   ``Status'' indicates changes since SW-1: ``old'' is
   unchanged, ``\texttt{new}'' is a new ensemble, and
    \texttt{update} indicates more measurements. 
      }
\label{tab:milc-lat}
\begin{ruledtabular}
\begin{tabular}{c | c | c | l | c | c }
$a$ (fm) & $am_\ell/am_s$ & size & ID & ens $\times$ meas & status \\
\hline
0.12 & 0.03/0.05  & $20^3 \times 64$ & C1 & $564 \times 9$   &  \texttt{update} \\
0.12 & 0.02/0.05  & $20^3 \times 64$ & C2 & $486 \times 9$   &  \texttt{update} \\
0.12 & 0.01/0.05  & $20^3 \times 64$ & C3 & $671 \times 9$   &  old \\
0.12 & 0.01/0.05  & $28^3 \times 64$ & C3-2 & $275 \times 8$ &  old \\
0.12 & 0.007/0.05 & $20^3 \times 64$ & C4 & $651 \times 10$  &  old \\
0.12 & 0.005/0.05 & $24^3 \times 64$ & C5 & $509 \times 9$   & \texttt{update} \\
\hline
0.09 & 0.0062/0.031 & $28^3 \times 96$ & F1 & $995 \times 9$ & \texttt{update} \\
\hline
0.06 & 0.0036/0.018 & $48^3 \times 144$ & S1 & $744 \times 2$ & old \\
\hline
0.045 & 0.0028/0.014 & $64^3 \times 192$ & U1 & $705 \times 1$ &\texttt{new}
\end{tabular}
\end{ruledtabular}
\end{table}

The sea-quark masses in these ensembles are not physical.
The strange quark is somewhat too heavy, requiring a small correction.
The light quarks are too heavy, requiring an extrapolation
to the physical mass. The lightest sea-quark pion on the coarse
ensembles has $m_\pi^{\rm min}({\rm sea}) \approx 280\;$MeV.
This will turn out to be light enough for a controlled extrapolation,
because the dependence on sea-quark masses turns out to be mild.

On each lattice, we create and destroy kaons using two wall-sources,
which are separated by Euclidean time $\Delta t$. The sources have the
property of creating only kaons with the desired Goldstone taste, $\xi_5$,
and with vanishing spatial momenta. The discretized version of
the operator $O_{\Delta S=2}$ is placed between the two sources, and
summed over space.
$\Delta t$ is chosen large enough that there is a plateau region
with minimal contamination from excited states, and small enough
that effects from states propagating ``around the world'' in the
time direction can be ignored. We obtain $B_K$ by fitting to a constant
over the plateau region. 
%
%

We use 10 different valence quark masses,
$a m_V= a m_s^{\rm nom} (n/10)$ for $n=1,2,\dots,10$,
with the nominal strange quark masses being
$0.05$, $0.03$, $0.018$ and $0.014$ for coarse, fine, superfine and
ultrafine lattices, respectively.
(For equal bare masses, HYP-smeared quarks have smaller
physical masses than asqtad quarks, because of differing
renormalization factors.)
%
%
We use the lightest four valence masses
for the valence $d$ quark ($m_x$), and the heaviest three valence
masses for the valence $s$ quark ($m_y$).
This maintains the relations 
$m_x \ll m_y \sim m_s^\text{phys}$, as required for
the applicability of SU(2) ChPT.

The four values of $a m_x$ allow us to extrapolate to
the physical down-quark mass.
This extrapolation is much shorter than that
for the sea-quarks, both since our valence quarks are lighter
(the lightest $\bar x x$ pion has a mass $\approx 200\;$MeV), 
and because we are extrapolating to $m_d^{\rm phys}$ 
rather than $(m_u^{\rm phys}\!+\!m_d^{\rm phys})/2$
(so that $M_{\bar x x}$ must be extrapolated to $158\;$MeV,
the mass of an unphysical $\bar d d$ meson).
We fit the dependence on $X=M_{\bar x x}^2$ 
(for pseudoscalar taste) to
the next-to-leading order (NLO) form predicted by
SU(2) SChPT. In SW-1 we used uncorrelated fits and
did not include finite-volume corrections.
Here we correct both shortcomings.
To obtain satisfactory correlated fits
(with $\chi^2/{\rm d.o.f.} \lesssim 1$) we need
to include higher-order terms with coefficients constrained
by Bayesian priors. Specifically, we fit $B_K$ for fixed $m_y$ to
\begin{equation}
c_1 F_0(X)\!+\!c_2 X\! +\! c_3 X^2 \!+\! c_4 X^2 \ln^2 X
\!+\! c_5 X^2 \ln X \!+\! c_6 X^3\,,
\end{equation}
where $F_0(X)$ contains the leading order constant term
as well as the chiral logarithms. The latter include taste-breaking
effects and finite-volume dependence (see SW-1 and
Ref.~\cite{FV11} for the explicit form).
The terms multiplied by $c_{3-5}$ are the
generic NNLO forms in the continuum.
Since these are not known analytically, we include them
with coefficients whose magnitude is constrained by Bayesian
priors to be of the size expected by naive dimensional analysis.
We also include a single analytic NNNLO term (with coefficient $c_6$).

\begin{figure}[t!]
  \includegraphics[width=20pc]{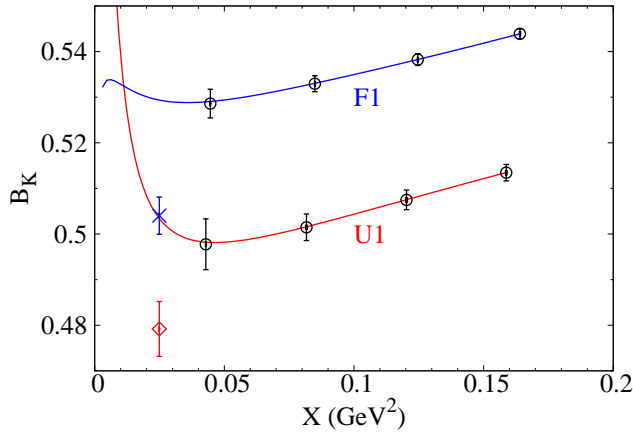}
  \caption{ $B_K({\rm NDR},1/a\;{\rm GeV})$ vs. $X$ on the U1 and
    F1 ensembles. The [red] diamond ([blue] cross) shows the result
    on the U1 (F1) ensemble,
    after extrapolation and removal of taste-breaking artifacts
    as described in the text.
  }
  \label{fig:bk-X-fit}
\end{figure}

Examples of the resulting fits are shown in Fig.~\ref{fig:bk-X-fit}.
With the fit parameters in hand, we can extrapolate
$m_x$ and $m_d^{\rm sea}$ to $m_d^{\rm phys}$, 
$m_u^{\rm sea}$ to $m_u^{\rm phys}$, the volume to infinity,
and remove taste-breaking discretization errors.
%
%
We note that finite volume shifts are small for our pion masses,
even though our lightest pions have $m_\pi L\approx 3$ (with $L$
the box size). We have checked this directly by comparing results on
ensembles C3 and C3-2~\cite{FV11}.

We estimate the systematic error of our X-fits by doubling
the allowed widths of the Bayesian priors, and by
dropping the NNNLO term. The former gives the larger effect,
and we take the largest size of this shift (0.33\%, on the S1
ensemble) as the error estimate.
%

The three values of $am_y$ allow us to extrapolate to the
physical strange quark mass.
%
%
We find that a linear fit to $m_y$ represents the data very well,
and use this for our central values. We use a quadratic ``Y-fit''
to estimate a fitting systematic.

\begin{figure}[t!]
  \includegraphics[width=20pc]{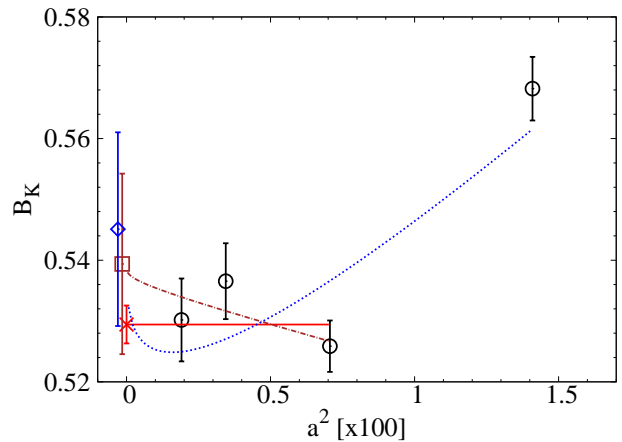}
  \caption{Continuum extrapolation of $B_K({\rm NDR},2\;{\rm GeV})$.
    The fits are described in the text.
     }
  \label{fig:scaling}
\end{figure}

At this stage the values of $B_K$ on different ensembles
differ primarily because of taste-conserving discretization
and matching errors. 
To remove the main part of these errors
we use ensembles C3, F1, S1 and U1, which have very similar
sea-quark masses.
In Fig.~\ref{fig:scaling}, we show the dependence on $a^2$,
and present the results of several methods of extrapolation to $a=0$.
We note that the simple linear dependence observed in SW-1 
(for the largest three lattice spacings) has been resolved
by improved errors, and by the addition of the U1 point,
into a less smooth dependence.
On theoretical grounds~\cite{VdWS}, the expected dependence is
\begin{equation}
d_1 + d_2 (a\Lambda)^2 + d_3 (a\Lambda)^2 
  \alpha_s + d_4 \alpha_s^2 + d_5 (a\Lambda)^4 + \dots
\,,
\label{eq:composite}
\end{equation}
where $\alpha_s=\alpha_s^{\overline{\rm MS}}(1/a)$.
We fit to this 5-parameter form applying Bayesian
constraints on $d_{2-5}$---the expected values are taken
to be $0$, while the standard deviations are set to $2$,
having chosen $\Lambda = 300\;$MeV.
%
%
The fit is shown by the [blue] dotted curve, and gives
the extrapolated value shown by the diamond. 
The fit, however, is very poor, with $\chi^2/\text{d.o.f.}=4.5$.
This problem is not resolved by adding terms of one higher order.
Thus we drop the coarse lattice from the fits,
and then find good fits to a constant (solid [red] line and cross),
a linear dependence on $a^2$ (not shown)
and to the constrained form (\ref{eq:composite}) 
(dashed [brown] line and square).
We take the constant fit for the central value, and the difference between it
and the constrained
fit as the systematic error in the continuum extrapolation.
For more discussion of fits see Ref.~\cite{wlee-2011-5}.

%
%

\begin{figure}[t!]
  \includegraphics[width=20pc]{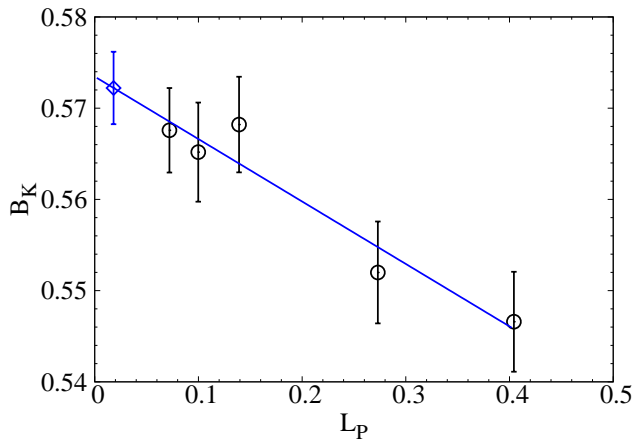}
  \caption{ $B_K({\rm NDR},2\;{\rm GeV})$ vs. $L_P$ on the coarse
    ensembles. 
    The solid line shows a linear fit,
    with the extrapolation to physical pion mass given by the diamond.
  }
  \label{fig:bk-sea-ml}
\end{figure}

After the preceding analysis, 
$B_K$ can still have a residual dependence on the sea quark masses.
In SU(2) SChPT, the dependence on $m_\ell$ is linear at NLO.
%
%
We have investigated the $m_\ell$ dependence 
in detail on the coarse lattices, with 
results shown in Fig.~\ref{fig:bk-sea-ml}.
We plot versus $L_P$, the squared mass of the sea-quark pion,
and find a linear behavior with a small slope
$\approx -1/(2.9\;{\rm GeV})^2$.
In SW-1, with errors 3 times larger, we could not uncover
this dependence.
Using this slope, we find that $B_K$ is increased by 1.5\%
when changing $L_P$ from its value on ensemble C3 to
its physical value. 
Since this is a small effect, and since we only
have results for the slope on the coarse ensembles,
we do not adjust the central value of $B_K$, but instead
quote the 1.5\% as a systematic error.

We also need to correct for the mismatch between 
$am_s$ and $a m_s^{\rm phys}$. Here we follow SW-1,
and assume the SU(3) ChPT form of the $m_s$ dependence.
Then we find that, on C3, the correction from using
a sea-quark mass which is 40\% too large is 1.3\%.
This approach is more conservative than that used in SW-1.
Again, we do not adjust the central value, but quote the
shift as a systematic error.

We collect all our errors in Table~\ref{tab:err-budget}.
%
%
The statistical error has been reduced by a factor of 3 compared to SW-1,
and now is smaller than many other sources of error.
The dominant error comes from our use of one-loop matching.
We estimate this as $\delta B_K/B_K=\alpha_s^2$,
with $\alpha_s$ evaluated at the scale of our finest lattice.
It is thus reduced from the 5.5\% estimate of SW-1 by the addition
of the U1 ensemble. 
The discretization error is nearly unchanged from SW-1,
despite the addition of the ultrafine lattices, because of the
increased uncertainty in the continuum extrapolation.
The only errors not discussed above are the
``$r_1$'' and ``$f_\pi$'' errors, which are estimated
essentially as discussed in SW-1.
%

\begin{table}[tbp]
\caption{Error budget for $B_K$ using SU(2) SChPT fitting.
  \label{tab:err-budget}}
\begin{ruledtabular}
\begin{tabular}{ l | l l }
  cause & error (\%) & memo \\
  \hline
  statistics       & 0.6   & see text \\
  matching factor  & 4.4   & $\Delta B_K^{(2)}$ (U1) \\
  discretization   & 1.9   & diff.~of constant and constrained fits \\
  X-fits           & 0.33  & varying Bayesian priors (S1) \\
  Y-fits           & 0.07  & diff.~of linear and quadratic (C3) \\
  $a m_l$ extrap   & 1.5   & diff.~of (C3) and linear extrap \\
  $a m_s$ extrap   & 1.3   & diff.~of (C3) and linear extrap \\
  finite volume    & 0.5   & diff.~of $V=\infty$ fit and FV fit\\
  $r_1$            & 0.14  & $r_1$ error propagation (C3) \\
  $f_\pi$          & 0.4   & $132\;$MeV vs. $124.4\;$MeV
\end{tabular}
\end{ruledtabular}
\end{table}
%
%
%

%

\begin{figure}[tbhp]
  \includegraphics[width=20pc]{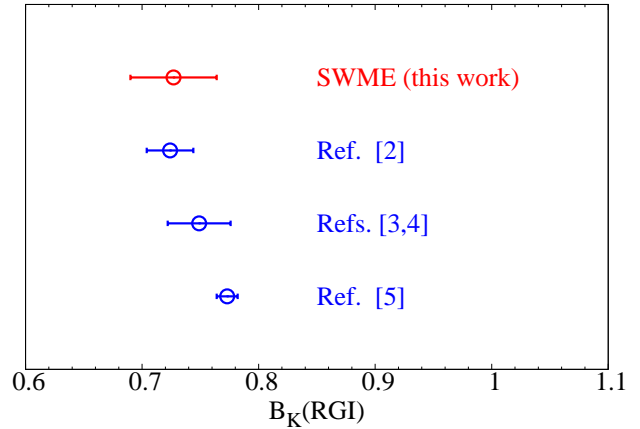}
  \caption{Comparison of our result 
for $\widehat{B}_K=B_K({\rm RGI})$ with other results obtained
with $N_f=2+1$ flavors.
}
  \label{fig:world}
\end{figure}

Adding  systematic errors in quadrature, we find
\[
\hat{B}_K = 
0.727 \pm 0.004(\text{stat}) \pm 0.038 (\text{sys})
%
%
\,.
\]
The central value is almost unchanged from SW-1, but the
significant improvements we have made have both reduced
the error and solidified our error estimates.
Our result is consistent with other $N_f=2+1$ 
results~\cite{ALV-09,rbc-uk-2010-1,BMW}, as shown in
Fig.~\ref{fig:world}.
The largest difference is that our result lies $~1.4\sigma$ below
that of Ref.~\cite{BMW} (the most accurate result). 
This consistency with results obtained using different fermion
discretizations is our most significant conclusion.
It is important, however, to further reduce errors in all lattice
calculations to check that this consistency holds up.
Work is in progress to reduce our dominant systematic using
two-loop matching and non-perturbative renormalization.

Lattice results for $B_K$ now allow the venerable experimental
result for $\epsilon_K$ to be used
to constrain the parameters of the SM. Indeed, we have now
reached the situation that the accuracy of $B_K$ calculations
is such that errors from other sources dominate---%
in particular those from $V_{cb}$ 
and the Wilson coefficient $\eta_{cc}$ (see, e.g., Ref.~\cite{Brod}).

\begin{acknowledgments}
We thank Claude Bernard for providing unpublished
information.
W.~Lee is supported by the Creative Research
Initiatives program (2012-0000241) of the NRF grant funded by the
Korean government (MEST).
C.~Jung and S.~Sharpe are supported in part by the US DOE 
through contract DE-AC02-98CH10886 and grant DE-FG02-96ER40956, respectively.
Computations for this work were carried out in part on the QCDOC computer 
of the USQCD Collaboration, funded by the Office of Science
of the US DOE.
W.~Lee acknowledges support from the KISTI supercomputing
center through the strategic support program [No. KSC-2011-C3-03].

\end{acknowledgments}

%

\bibliographystyle{apsrev} 
\bibliography{ref} 

\end{document}